\documentclass[final,1p,times]{elsarticle}
\usepackage{multirow}
\usepackage{amssymb}
\usepackage{color}
\usepackage{graphicx}% Include figure files
\usepackage{dcolumn}% Align table columns on decimal point
\usepackage{bm}% bold math
\usepackage{array}
\usepackage{subfig}

\newcommand{\PreserveBackslash}[1]{\let\temp=\\#1\let\\=\temp}
\newcolumntype{C}[1]{>{\PreserveBackslash\centering}p{#1}}
\newcolumntype{R}[1]{>{\PreserveBackslash\raggedleft}p{#1}}
\newcolumntype{L}[1]{>{\PreserveBackslash\raggedright}p{#1}}
\journal{Physics A}

\begin{document}
\begin{frontmatter}

\title{Improving personalized link prediction by hybrid diffusion}

\author[inst1]{Jin-Hu Liu}
\author[inst1]{Yu-Xiao Zhu\corref{cor1}}
\author[inst1,inst2]{Tao Zhou}
\cortext[cor1]{zhuyuxiao.mail@gmail.com(Yu-Xiao Zhu)}
\address[inst1]{Complex Lab, Web Sciences Center, University of Electronic Science and Technology of China, Chengdu 611731, China}
\address[inst2]{Big Data Research Center, University of Electronic Science and Technology of China, Chengdu 611731, China}

\begin{abstract}
Inspired by traditional link prediction and to solve the problem of recommending friends in social networks, we introduce the personalized link prediction in this paper, in which each individual will get equal number of diversiform predictions. While the performances of many classical algorithms are not satisfactory under this framework, thus new algorithms are in urgent need. Motivated by previous researches in other fields, we generalize heat conduction process to the framework of personalized link prediction and find that this method outperforms many classical similarity-based algorithms, especially in the performance of diversity. In addition, we demonstrate that adding one ground node who is supposed to connect all the nodes in the system will greatly benefit the performance of heat conduction. Finally, better hybrid algorithms composed of local random walk and heat conduction have been proposed. Numerical results show that the hybrid algorithms can outperform other algorithms simultaneously in all four adopted metrics: AUC, precision, recall and hamming distance. In a word, this work may shed some light on the in-depth understanding of the effect of physical processes in personalized link prediction.

\end{abstract}


\end{frontmatter}

\section{Introduction}

Network has been used as a useful model to describe many social, biological and information systems, where nodes represent individuals and links reflect the relations or interactions between nodes~\cite{newman_structure_2003,Boccaletti2006,Cohen2010}. Networks have been widely studied in many different fields
and one of fundamental problems for network analysis is link prediction, which
aims to estimate the likelihood of the existence of a link between two nodes based on observed links and the attributes of nodes~\cite{survey-LP,linkmining}. For example, the existence of a link must be verified by costly chemical experiments
in many biological networks, such as protein-protein interaction networks and metabolic networks. If the predictions are accurate enough, the experimental cost can be sharply reduced compared to blindly checking. Missing data problem is also exist in social network, where link prediction is also one useful tool.
In addition, link prediction algorithms can also be applied to identify spurious links~\cite{spurious1,spurious3,spurious4}.
Link prediction algorithms can not only be used to predict missing data but also practical to predict the links that may appear in the future of evolving networks. For example, in online social networks, very likely but not yet existent links can be
recommended as promising friendships, which can help users to find new friends and thus enhance their loyalties to the websites.

In the traditional link prediction, all the nonexistent links are sorted in descending order according to their prediction scores, and the top-ranked links are most likely to exist. Clearly, in this case, the prediction list is generated from a global perspective, in which some nodes may have large number of promising links while others may have very few or even zero possible links. This straightforward and standard method may lead to some bias. On the one hand,
in this case, the links connect low-degree nodes may be ignored casually, while this kind of information may very be important and meaningful~\cite{zhu_2012}. In addition, some research unveiled that low-degree users may have a big influence in the future~\cite{lowdegree}. On the other hand, the imbalance of prediction list may bring unsatisfactory for some individuals and thus affect the experience of the whole system. For example, in social networks, accurately predicting certain number of potential friends or acquaintances for each registered user is useful and meaningful. In this case, no real distinction can be made between low-degree and high-degree users and global link prediction
does not apply in this case decently. However, this phenomenon has always been neglected in the traditional link prediction for the past several decades. To solve these problems, we propose personalized link prediction here, in which all nodes will get equal number of possible links through their own past link records.

One challenge deserved special attention recently, called low-diversity problem, has plagued almost all recommendation systems. It means that lots of recommender systems always recommend very similar items to different users which narrows users' views~\cite{HC}. Subsequently, some physical dynamics, like heat conduction process (HC) have been applied to design recommender systems and can improve the diversity of recommendation. Motivated by this, we generalize heat conduction process to the framework of personalized link prediction and find that it outperforms other methods in diversity but
do not perform very satisfactorily in accuracy. To solve this dilemma, ground node, who is supposed to connect all the nodes in the system, is incorporated to improve the prediction accuracy. Finally, we generalize one superior hybrid algorithm (LH) and propose another better hybrid algorithm (LGH) composed of local random walk (LRW)~\cite{lrw} and ground heat conduction (GHC), which performs pretty well not only on accuracy but also on diversity.

This article is organized as follows. In the next section, we will clearly define the problem of personalized link prediction,
describe the standard metrics for evaluation. Then we explain several state-of-the-art similarity indices and introduce new algorithms HC, GHC, LH, LGH in Section 3. Data description and experimental results for the existed predicting algorithms and the proposed method are presented in Section 4.
Finally, we summarize our results in Section 5.

\section{Problem and Metrics}
%\subsection{personalized link prediction}
For one given undirected network $G(V,E)$, in which $V$ and $E$ are the sets of nodes and links respectively. The universal set of all $\frac{|V|(|V|-1)}{2}$ possible links are denoted by $U$, where $|V|$ denotes the number of elements in set $V$ (multiple links and self-connections are not allowed). Clearly the set of nonexistent links is $U\setminus E$, in which there are some missing links (i.e.,the existed yet unknown links) and promising links (i.e.,very likely but not yet existent links). The task of link prediction is to uncover these links. Each node pair $x$ and $y$ will be assigned a scores $s_{xy}$ according to a given prediction algorithm.
The higher the score is, the higher existence likelihood this link has. For each node $x$,
we denote the set of its \emph{revelent nonexistent links} (nonexistent links that connect
$x$) as $(U\setminus E)_{x}$, thus all links in $(U\setminus E)_{x}$ are sorted
in descending order according to their scores, and the top-ranked links are most likely to exist.

To test the performance of one given algorithm, we divide the observed links $E$ into two sets: the training set $E^{T}$ (considered as known information) and the test set $E^{P}$ (used for testing and no information therein is allowed to be used for prediction). Clearly, $E =E^{T}\cup E^{P}$ and $E^{T}\cap E^{P}=\phi$. For each node $x$, the relevant test set (links in $E^{P}$ that connect
$x$) is denoted by $E_{x}^{P}$. We then introduce four popular evaluation
metrics as below.

%In traditional link prediction problem, all the nonexistent links are sorted
%in descending order according to their scores, and the top-ranked links are most likely to exist~\cite{survey-LP}.
%While in this paper, we consider personalized link prediction problem, in which generate top-ranked links for each node
%separately. That is to say, each node will get one predicted list accordingly.

%\subsection{Metrics}
(i) AUC - short for area under the receiver operating characteristic curve, is considered as one standard metric to quantify the accuracy of prediction~\cite{AUC}.
Specifically, for each node $x$, this metric can be interpreted as the probability that a randomly chosen revelent relevant missing link (links in $E^{P}_{x}$) has higher score than a randomly chosen relevant nonexistent link (links in $(U\backslash E)_{x}$). In the implementation, among $n$ times of independent comparisons, if there are $n_{1}$ times that the missing link has higher score and $n_{2}$ times the missing link and nonexistent link have the same score, the AUC value is defined as
\begin{equation}
AUC=\frac{n_{1}+0.5n_{2}}{n}.
\end{equation}
The AUC of the whole system is the average value over all nodes in the system.
If all the scores are generated from an independent and identical distribution, the accuracy should be about 0.5. Therefore, the extent to which the accuracy exceeds 0.5 indicates how much better the algorithm performs than pure chance.

(ii) Precision and Recall \cite{survey-LP} - Given the ranking of the non-observed links, the precision is defined as the ratio of relevant items selected to the number of items selected.
Denoting by $L$ the length of prediction list (i.e. the number of nodes recommended to each individual). For each individual $x$, if we take the top-$L$ links
as the predicted ones, among which $L_{x}$ links are right (i.e., there are $L_{x}$ links in the test set $E^{P}_{x}$), then the precision equals $P_{x} = L_{x}/L$.
While recall is defined as the ratio of relevant items selected to the number
of relevant items in the testing set. That's, $R_{x} = L_{x}/N_{x}$, where
$N_{x}$ denotes the number of node $x$'s positive edges in its testing set
$E^{P}_{x}$. Clearly, higher precision and higher recall means higher prediction accuracy. The precision (recall) of the whole system can be calculated by the average value among all individuals.

(iii) Hamming distance \cite{hamming,zhou_NJP} - One of the famous metrics that quantify the intra-diversity
of the prediction system. For individual $x$ and $y$, if the overlapped number of nodes in $x$ and $y$'s prediction lists is $Q_{xy}$, their Hamming distance is defined as
\begin{equation}
H_{xy} = 1-\frac{Q_{xy}}{L}.
\end{equation}
Generally speaking, a more diverse prediction list should have larger Hamming
distances which means recommending appropriately but not popularly. Accordingly, we use the mean value of Hamming distance,
\begin{equation}
H = \frac{1}{|V|(|V|-1)}\sum_{x\neq y}H_{xy},
\end{equation}
averaged over all the node-node pairs, to measure the diversity of predictions.

\section{Algorithms}

\subsection{similarity-based algorithms}

In the traditional link prediction problem, the study on similarity-based algorithms is the mainstream due to its simplicity. Considering this, we adopt the simplest local similarity indices as benchmark in the framework of
personalized link prediction in this paper. For similarity-based algorithm, the aforementioned scores
$s_{xy}$ is directly defined as the similarity between node $x$ and $y$~\cite{Kleinberg2007}.
For each node $x$, rank all \emph{relevant links (links connect $x$ and other nodes)}
in relevant non-observed set $(U\setminus E)_{x}$ based on their scores, and
links with higher scores are supposed to be of higher existence likelihoods
and thus regarded as personalized prediction list (we consider length of list $L=5, 20$).
We will compare the performances of these indices on personalized link prediction and
the details of these indices are as follows.

(1) Common Neighbors (CN).
In common sense, two nodes $x$ and $y$ are more likely to have a link if they have many common neighbors~\cite{CN}.
The simplest measure of its neighborhood overlap is the direct count, namely
\begin{equation}
s^{CN}_{xy}=|\Gamma_{x} \cap \Gamma_{y}|,
\end{equation}
in which $\Gamma_{x}$ denote the set of neighbors of $x$.

(2) Salton index~\cite{Salton}. It is defined as
\begin{equation}
s^{Salton}_{xy}= \frac{|\Gamma_{x} \cap \Gamma_{y}|}{\sqrt{k_{x}k_{y}}},
\end{equation}
where $k_{x}$ denotes the degree of node $x$. The Salton index is also called the cosine similarity in the literature.

(3) Jaccard index~\cite{Jaccard}. This index was proposed by Jaccard over one hundred years ago, defined as
\begin{equation}
s^{Jaccard}_{xy}=\frac{|\Gamma_{x} \cap \Gamma_{y}|}{|\Gamma_{x} \cup \Gamma_{y}|}.
\end{equation}

(4) Adamic Adar (AA) index~\cite{AA}. This index refines the simple counting of common neighbors by assigning the less-connected neighbors more weights, and is defined as
\begin{equation}
s^{AA}_{xy} = \sum_{z\in \Gamma_{x} \cap \Gamma_{y}}\frac{1}{log k_{z}}.
\end{equation}

(5) Preferential Attachment (PA)~\cite{PA}. The mechanism of preferential attachment is widely used
to generate evolving scale-free networks, where the probability that a new link is connected to
the node $x$ is proportional to $k_{x}$. The corresponding similarity index can be defined as
\begin{equation}
s^{PA}_{xy} = k_{x} \times k_{y}.
\end{equation}

(6) Resource Allocation (RA) index~\cite{RA}. This index is motivated by the resource allocation dynamics on complex
networks, and is defined as
\begin{equation}
s^{RA}_{xy} = \sum_{z\in \Gamma_{x} \cap \Gamma_{y}}\frac{1}{k_{z}},
\end{equation}
where $z$ runs over all common neighbors of $x$ and $y$.

(7) Local Random Walk (LRW) index~\cite{lrw}. To measure the similarity between nodes $x$ and $y$, a random walker is initially put on node $x$ and thus the initial density vector $\overrightarrow{\pi_{x}}(0)=\overrightarrow{e_{x}}$. This density vector evolves as $\overrightarrow{\pi_{x}}(t+1)=P^{T}\overrightarrow{\pi_{x}}(t)$ for $t\geq0$. The LRW index at time step $t$ is thus defined as
\begin{equation}
s^{LRW}_{xy}(t) = q_{x}\pi_{xy}(t) + q_{y}\pi_{yx}(t),
\end{equation}
where $q$ is the initial configuration function.

\subsection{Heats Conduction algorithm and Hybrid algorithm}

\begin{figure*}[htpb]
  \centering
  \includegraphics[trim=2mm 2mm 2mm 2mm, width = 13cm,height=8cm]{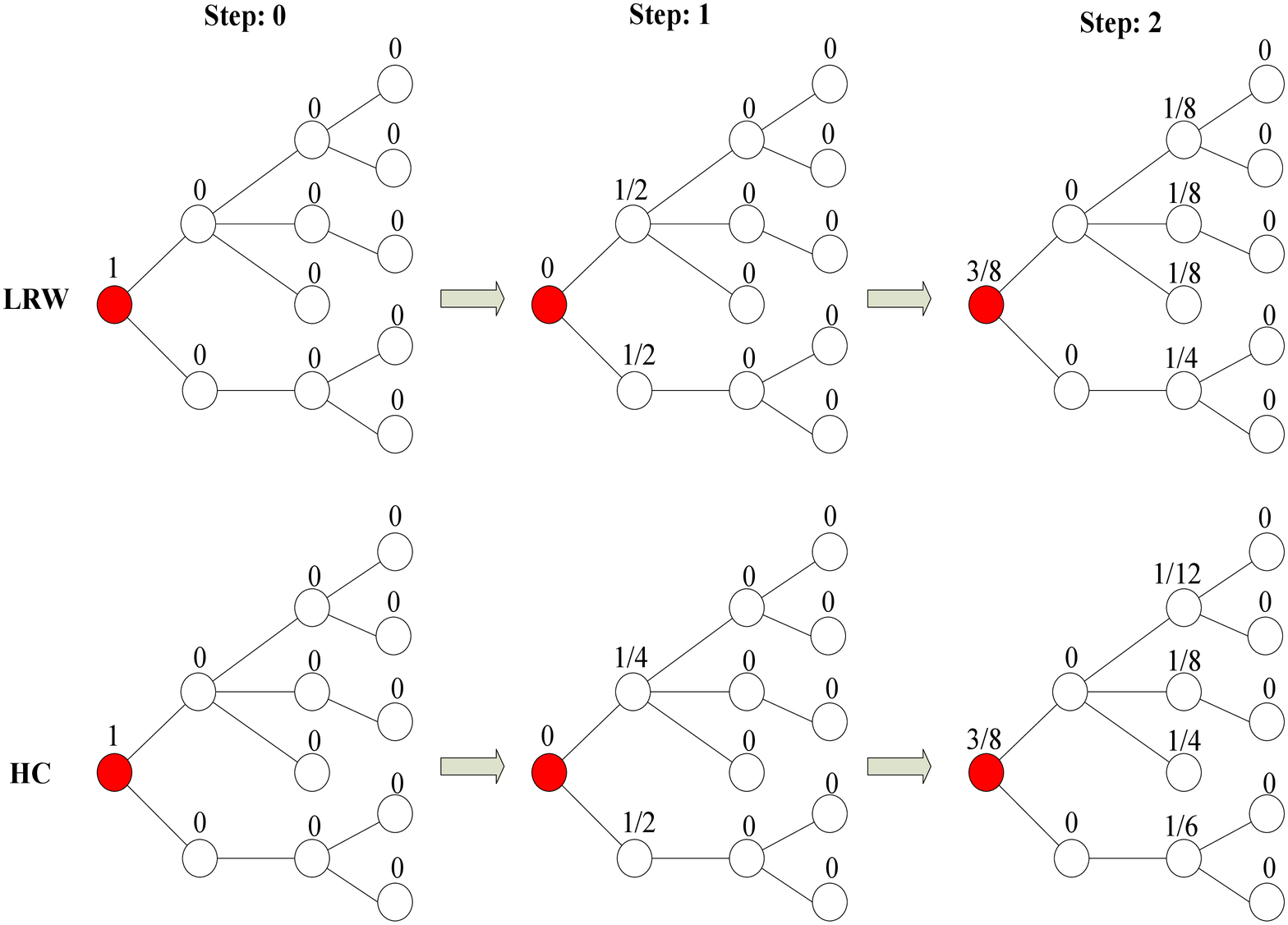}
  \caption{The schematic drawing that illustrates LRW and HC from Step:0 to Step:2. Nodes are represented by circles and lines represent the existent links between them. The numerical value over each node indicates its temperature resource in this step. The red node is the target node. In each step after step 2, those high-temperature nodes which have no links with the target node will be recommended. }\label{demo}
\end{figure*}

Motivated by some previous studies~\cite{HC}, here we generalize the heat conduction algorithm to the framework of personalized link prediction. Basically,
heats conduction algorithm (HC) recommends promising links to an individual node
through a process motivated by heat diffusion. Firstly, the adjacency matrix, denoted by $A$, where the element $a_{xy}=1$ if node $x$ has connection with node $y$, $a_{xy}=0$ otherwise.
Let assume one target node $x$ here, the temperature resource
for $x$ is initialized with 1, while 0 for the remaining nodes.
Then in each iteration, the HC algorithm redistributes
the temperature resource via a nearest-neighbor averaging process.
That is to say, the temperature resource of node $x$ in $(t+1)$th iteration can be written as:
\begin{equation}
T_{x}(t+1)= \frac{1}{k_{x}}\sum_{y=1}^{M}a_{xy}T_{y}(t),
\label{HC}
\end{equation}
where $M$ denotes the number of nodes. The degree of node $x$, the number of nodes who have connections with $x$, is denoted as $k_{x}$.
Then the final temperatures after the diffusion are considered as
the corresponding scores and the resulting top-ranked list of non-connected nodes
is sorted according to these scores in descending order. In order to describe the detailed diffusion process visually, an illustration of the first two step of HC and LRW processes is shown in Figure.~\ref{demo}.

In depth, we also propose a new algorithm by adding a ground node who connect all the nodes in the network~\cite{lv-groundnode}. The iterative temperature of the GHC (abbreviation of the HC with ground node) algorithm is thus written as
\begin{equation}
T_{x}(t+1)= \frac{1}{k_{x}+1}(\sum_{y=1}^{M}a_{xy}T_{y}(t)+T_g(t)).
\label{GHC}
\end{equation}
where $T_g$ denotes the temperature of the ground node, which leads to an additional link between two nodes
even when they don't have connection.
%Here we will show that the ground node also benefits the personalized link prediction.
%Experimental results show
%that by adding the ground user the recommendation accuracy will
%be increased.
%The recommendation is made according to the
%equilibrium temperature of the nodes in the networks.

In addition, motivated by the literatures~\cite{HC,LRR}, we generalize one superior hybrid algorithm (LH) composed of LRW and HC.
\begin{equation}
T_{x}(t+1)= \alpha \sum_{y=1}^{M} \frac{a_{xy}T_{y}(t)}{k_y}+\frac{1-\alpha}{k_{x}}\sum_{y=1}^{M}a_{xy}T_{y}(t).
\label{LH}
\end{equation}
where $\alpha$ is an adjustable parameter which ranges from 0 to 1. Obviously, LH turns to HC when $\alpha=0$, while degenerates to
LRW when $\alpha=1$. This method can well test potential nodes from two aspects, one is the strength of joint and the other is the personalization. Furthermore, we propose another novel hybrid algorithm (LGH) by combining LRW and GHC.
\begin{equation}
T_{x}(t+1)= \alpha \sum_{y=1}^{M} \frac{a_{xy}T_{y}(t)}{k_y}+\frac{1-\alpha}{k_{x}+1}(\sum_{y=1}^{M}a_{xy}T_{y}(t)+T_g(t)).
\label{LGHC}
\end{equation}
where $\alpha$ is an adjustable parameter which ranges from 0 to 1. Obviously, LGH turns to GHC when $\alpha=0$, while degenerates to
LRW when $\alpha=1$.

\section{Data and Experiments}
\subsection{Data}
We consider four representative networks drawn from disparate fields: (i) USAir. The network of the USAir transportation system, which contains 332 airports and 2126 airlines~\cite{USAir}.
(ii) C.elegans (CE). The neural network of the nematode worm C.elegans, in which an edge joins two neurons if they are connected by either a synapse or a gap junction~\cite{CE}. This network contains 297 neurons and 2148 links.
(iii) Political Blogs (PB). The network of US political blogs~\cite{PB}. The original links are directed, here we treat them as undirected links.
(iv) Food Webs (FW). The food circle network of Florida bay, containing 128 living beings and 2106 preying relations.
Table.~\ref{data} summarizes the basic topological features of these networks~\cite{FW}.
Brief definitions of the monitored topological measures can be found in the table caption.

%%%%%%%%%%%%%%%% Table 1 %%%%%%%%%%%%%%%%%
\begin{table*}[htpb]
  \centering
  \caption{The basic topological features of the giant components of the four example networks.
  CE, PB and FW are the abbreviations for C.elegans, Political Blogs and Food Webs networks, respectively.
  $N=|V|$ and $M=|E|$ are the total number of nodes and links, respectively.
  $C$ is the clustering coefficient that is defined as the average ratios of the number
  of connected pairs of a node's neighbors to the possible maximum~\cite{cc}.
  $r$ is the assortative coefficient~\cite{assortative}, the Pearson correlation coefficient of degree
  between pairs of connected nodes. $r$ lies between $-1$ and $1$,
  and $r>0$ indicates a positive correlation while $r<0$ indicates a negative correlation.
  $\langle k\rangle$ is the average degree of the network, and $\langle d\rangle$ is the
  average shortest distance between node pairs. $H$ denotes the degree heterogeneity defined
  as $H=\frac{\langle k^{2} \rangle}{\langle k \rangle^{2}}$.
  }
  {\begin{tabular}{cccccccccccccccccccccc}
  \hline\hline Datasets & $N$ & $M$ & $C$ &  $r$ & $\langle k \rangle$ & $\langle d \rangle$ &H \\
  \hline
  \textbf{USAir}& 322&  2126& 0.749& -0.208& 12.81 & 2.46 & 3.46\\
  \textbf{CE}& 297&  2148& 0.308& -0.163& 14.46 & 2.46 & 1.80\\
  \textbf{PB}& 1222&  16717& 0.361& -0.221& 27.36 & 2.51 & 2.97\\
  \textbf{FW}& 128 & 2106 & 0.335 & -0.104 & 32.90 & 1.77 & 1.23 \\
  \hline
  \end{tabular}
  \label{data}}
\end{table*}

 % \textbf{NetScience}& 379& 914& 0.798& -0.082& 4.82 & 4.93 & 1.66\\
   %\textbf{Yeast}& 2375 & 11693 & 0.388 & 0.454 & 9.85 & 4.59 & 3.47 \\

\subsection{Experimental Results}

To test the performances of algorithms, we
divide the data sets into training set and testing set randomly in our experiments. The ratio of training set to testing set is 9:1, that is to say,
testing set contains 10\% links. The prediction list for each individual is provided based on the training set, and the testing set will be used for testing. Four metrics like AUC, precision, recall and hamming distance
are adopted here to give quantitative measurements of the methods. All the results below are averaged over 100
independent runs with different data divisions.

%%%%%%%%%%%%%%% Table 1 %%%%%%%%%%%%%%%%%
\begin{table*}[htpb]
  \centering
  \caption{Algorithmic accuracies on four different datasets, measured by AUC.
  Each value is obtained by averaging over 100 implementations with independently divisions of training set
  and test set randomly. CE, PB and FW are the abbreviations for C.elegans, Political Blogs and Food webs networks respectively. The number in bracket indicates the step in which the corresponding algorithm gets optimal value.}
  {\begin{tabular}{ccccccccccc}
  \hline\hline Datasets & USAir & CE & PB & FW  \\
  \hline
  \textbf{CN}      & 0.9354 & 0.8285 & 0.8645 & 0.5516  \\
  \textbf{Salton}  & 0.9049 & 0.7979 & 0.8129 & 0.4988  \\
  \textbf{Jaccard} & 0.8853 & 0.7933 & 0.8018 & 0.4988  \\
  \textbf{AA}      & 0.9432 & 0.8405 & 0.8688 & 0.5589  \\
  \textbf{PA}      & 0.8945 & 0.7724 & 0.8499 & 0.6771  \\
  \textbf{RA}      & 0.8976 & 0.8440 & 0.8697 & 0.5656  \\
  \textbf{LRW}     & 0.9409(4) & 0.8821(3) & 0.9153(3) & 0.8412(3) \\
  \textbf{HC}      & 0.8450(3) & 0.8548(3) & 0.8426(3) & 0.8804(3) \\
  \textbf{GHC}     & 0.8697(3) & 0.8614(3) & 0.8651(3) & 0.8821(3)  \\
  \textbf{LH}     & 0.9507(4) & 0.8961(3) & 0.9189(3) & 0.9086(3)  \\
  \textbf{LGH}    & \textbf{0.9539(4)} & \textbf{0.8973(3)} & \textbf{0.9204(3)} & \textbf{0.9091(3)} \\
  \hline
  \end{tabular}
  \label{AUC}}
\end{table*}

\begin{figure*}[htpb]
  \centering
  \includegraphics[trim=20mm 20mm 20mm 20mm, width = 12cm,height=9cm]{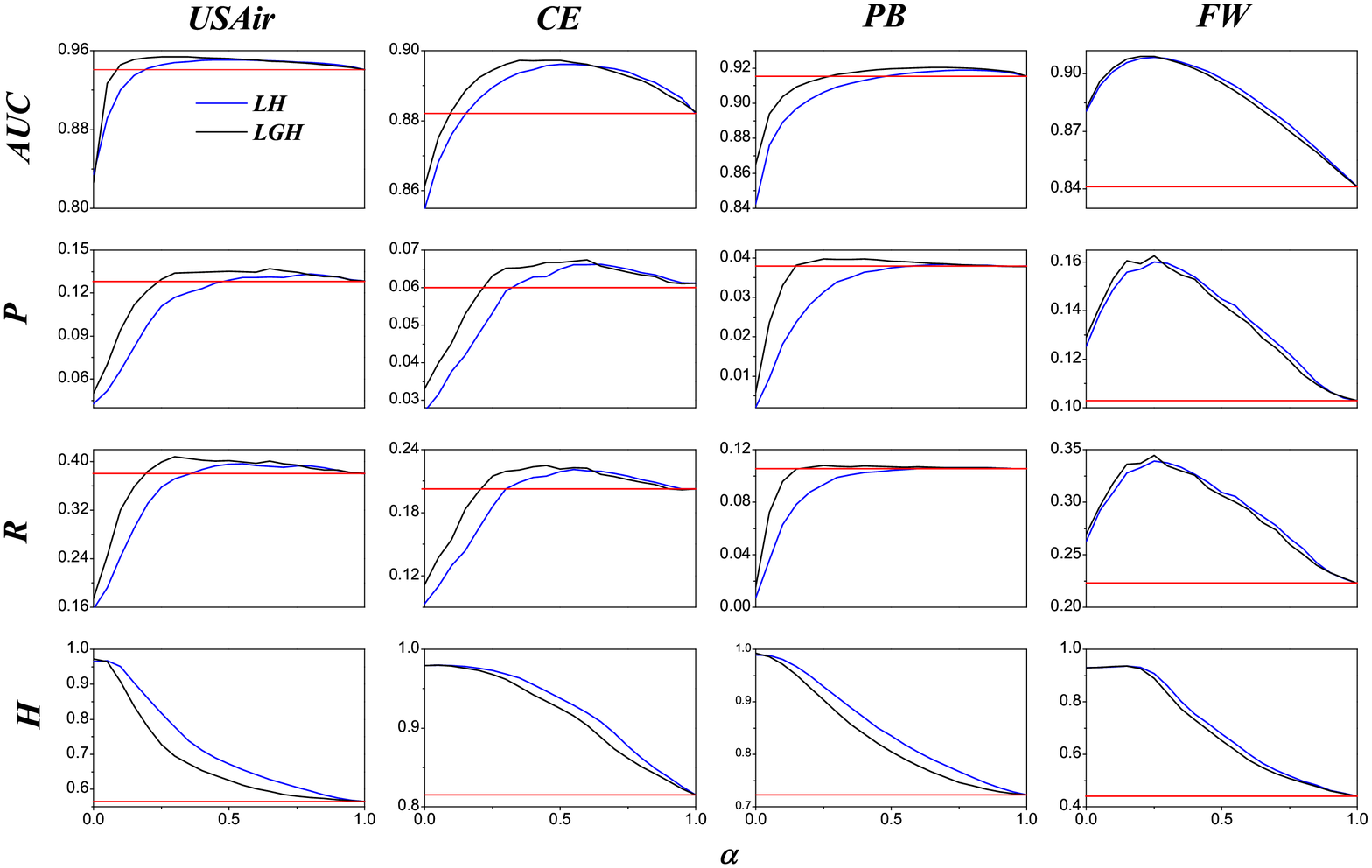}
  \caption{The performance of LH and LGH measured by four
   metrics (AUC, precision (P), recall (R) and hamming distance (H)) as
   a function of $\alpha$ in the representative data sets. The blue and black curves represent results of LH and LGH in the step getting optimal AUC, respectively. And the red solid line represents results of LRW in the step getting optimal AUC.
   All the numerical results are obtained by averaging over 100 independent runs with data division. We set $L=5$ here.}\label{L5}
\end{figure*}

\begin{figure*}[htpb]
  \centering
  \includegraphics[trim=20mm 20mm 20mm 20mm, width = 12cm,height=9cm]{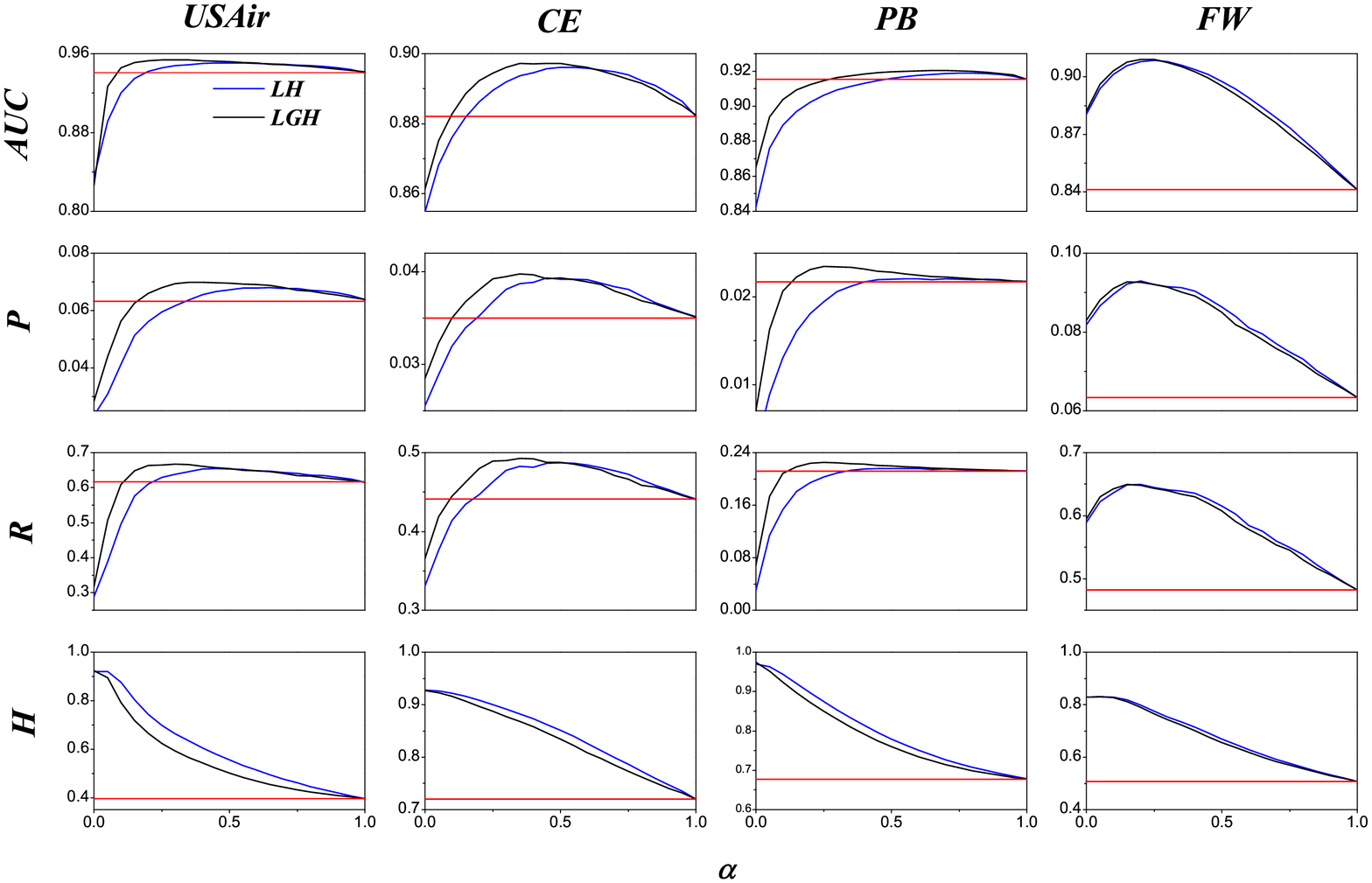}
  \caption{The performance of LH and LGH measured by four
   metrics (auc, precision (P), recall (R) and hamming distance (H)) as
  a function of $\alpha$ in the representative data sets. The blue and black curves represent results of LH and LGH in the step getting optimal AUC, respectively. And the red solid line represents results of LRW in the step getting optimal AUC.
  All the numerical results are obtained by averaging over 100 independent runs with data division identical to the case shown in Figure~\ref{L5}. We set $L=20$ here.
  } \label{L20}
\end{figure*}

The performances of different methods measured by AUC in all data sets
are shown in Table.~\ref{AUC}. Highest AUC value in each column is emphasized in black. The optimal step of iterations correspond to best AUC are shown in the brackets. GHC is an abbreviation of the method HC with a ground node. LH refers to the hybrid method that combines LRW and HC algorithms, while LGH refers to the hybrid method that combines LRW and GHC algorithms.
Clearly, HC algorithm outperforms classical benchmark algorithms (CN, Salton, Jaccard, AA and PA) in most networks, while is slightly inferior to LRW. In order to improve the prediction accuracy of HC, we propose one new method called GHC by adding one ground node in the system. By comparing the results of HC and GHC, we can see that the prediction accuracy can be improved to some extent by this way. For example, the AUC increases from 0.8450 to 0.8697 for USAir data set. Previous studies have shown that the original HC algorithm prefers to the small-degree nodes~\cite{HC}. Adding one ground node to the system practically amounts to add an additional transition probability from ground node to another. So in every iteration,
each node receives the same heat from the ground node and then average it with the heat from other sources, thus the temperature of the big-degree nodes will be enhanced. This improvement of the accurate predictions on big-degree nodes leads to the improvement of prediction accuracy of the whole system. Moreover, diffusion-based algorithms will be restricted by the network connectedness. The ground node happens to make the whole network much more strongly connected and the shortest path between any two nodes is less than 3. That's why GHC performs better than HC.

\begin{figure*}[htpb]
  \centering
  \includegraphics[trim=20mm 0mm 20mm 10mm, width = 12cm,height=8cm]{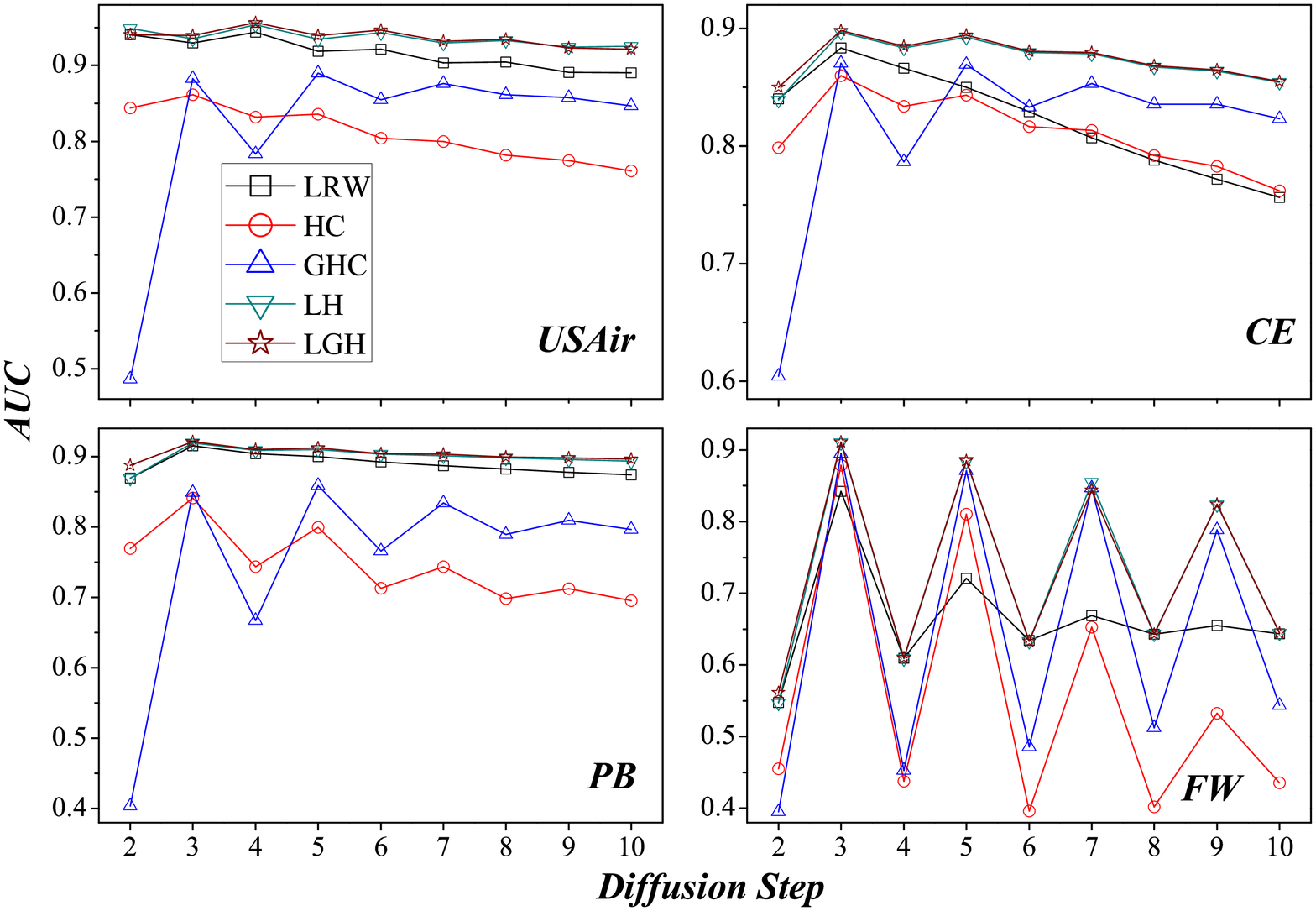}
  \caption{The correlation between AUC and the diffusion step. The black squares, red circles, blue triangles, cyan triangles and wine stars represent the AUC of the five algorithms in the each diffusion step, respectively. The values of AUC fo LH and LGH are all corresponding to optimal values in each step.}\label{Steps}
\end{figure*}

Furthermore, in Figure.~\ref{L5}, the black and blue curves show the results of LGH and LH with the corresponding $\alpha$ respectively, for comparison, results of LRW ($\alpha=1$) are displayed in red solid line. Obviously, hybrid algorithms have greatly improved the prediction results. By combining LRW and GHC, the hybrid algorithm (LGH) performs the best in the accuracy of prediction measured by AUC. Besides, Figure.~\ref{L5} displays the performances of LGH measured by recall and precision in relationship with free parameter $\alpha$, although its diversity is a little lower than LH. The advantage of hybrid algorithms comes from two aspects. Firstly, LRW makes those popular nodes still having determinate weighting to exist in the prediction list. Secondly, more importantly, GHC or HC enhances the weights of those low-degree nodes and reduces the weights of high-degree nodes simultaneously, which make the prediction lists much more personalized and accurate. The hybrid algorithms with certain range of free parameter
not only improve the accuracy (AUC, precision, recall) but also befit
the diversity of prediction. The improvements of this hybrid algorithms
are independent of prediction length $L$. Figure.~\ref{L20} shows the corresponding
results when $L=20$, which display the same tendency.

For all diffusion-based method, one key problem is the diffusion step. {Liu et al. have already proven a positive correlation between the optimal step and the average shortest distance~\cite{lrw}. In our algorithm, the average shortest distance of a network is getting smaller with the adding of ground node. So our algorithm can quickly obtain the optimal value with less diffusion steps. From Table. \ref{AUC} and Figure. \ref{Steps}, it is clearly that LH and LGH both obtain their best results within four steps in all the four data sets. And after the 5th step, the results in all data sets get worse with the increase of diffusion step.

\section{Conclusion and Discussion}

In the past several decades, Internet is flourishing as never before. Especially, as new free-registered platforms, social networks provide us with abundant information, countless items and infinite opportunities to make fresh friends around the world. Meanwhile, we are in deep trouble when facing information overload. We are extremely hard to dig out interesting information, suitable items to purchase, new friends with the same abilities and interests and old acquaintances we lost touch with. Fortunately, the appearance of personalized recommendation and prediction gives us considerable assistance. Based on historical behaviors, potential interests and links will be found out automatically via appropriate algorithms. Thus, an accurate and effective algorithm is vital and extremely valuable. But to our knowledge, in the previous studies on link prediction,
all the nonexistent links are sorted in descending order according to their prediction scores, and the top-ranked links are most likely to exist. Clearly, in this case, the top-ranked list is generated from a global perspective, in which some nodes may have large number of promising links while others may have very few or even zero possible links. Considering this bias, we propose personalized link prediction problem in this paper, in which all nodes will get equal number of possible links. Motivated by some previous works, we generalized the physical process - heat conduction to the framework of personalized link prediction and find that it outperforms most existing similarity-based algorithms. In addition, we demonstrate that adding one ground node which is supposed to connect all the nodes in the system will benefits the performance of HC. Finally, we introduce two hybrid algorithms that perform pretty well not only on accuracy but also diversity. Note that, one small defect of our algorithms is relatively complicated and time-cost, fortunately this weakness can be easily solved by parallel computation technique. With the improvement of calculating technique, an algorithm with better performance must be a top choice.

However, how to provide better personalized predictions is still a long-standing challenge in
modern information science. One satisfying answer to this question may benefits
our society, economic and lifestyle in the near future. As a
starting point, we give a naive method and a preliminary analysis, which is of course far from a satisfactory answer to the question. In fact,
we believe the current work can provide some insights to understand this issue.

\section{Acknowledgments}
This work was partially supported by the National Natural Science Foundation of China under grant nos. 61370150 and 61433014, and the Program of Outstanding
PhD Candidate in Academic Research by UESTC: Y-BXSZC20131035. The funders had no role in study design, data collection and analysis, decision to publish, or preparation of the manuscript.

\bibliography{bibfile}

\end{document}